\documentclass[conference]{IEEEtran}
\IEEEoverridecommandlockouts
\usepackage{cite}
\usepackage{amsmath,amssymb,amsfonts}
\usepackage{algorithmic}
\usepackage{graphicx}
\usepackage{textcomp}
\usepackage{xcolor}
\usepackage{subcaption}
\usepackage{multirow}
\usepackage{xfrac}
\usepackage{nicefrac}

\def\BibTeX{{\rm B\kern-.05em{\sc i\kern-.025em b}\kern-.08em
    T\kern-.1667em\lower.7ex\hbox{E}\kern-.125emX}}
\begin{document}

\title{ Sentiment Analysis of Microblogging dataset on Coronavirus Pandemic}

\author{\IEEEauthorblockN{Nosin Ibna Mahbub}
\IEEEauthorblockA{\textit{Dept. of Information and}\\ \textit{Communication Technology,} \\
\textit{Islamic University}\\
Kushtia-7003, Bangladesh.\\
mahbub.nosin@gmail.com}
\and

\IEEEauthorblockN{Md. Rakibul	Islam}
\IEEEauthorblockA{\textit{Dept. of Information and}\\ \textit{Communication Technology} \\
\textit{Islamic University}\\
Kushtia-7003, Bangladesh.\\
mrpislam@gmail.com}
\and

\IEEEauthorblockN{Md. Al Amin}
\IEEEauthorblockA{\textit{Dept. of  Computer} \\
\textit{Science \& Engineering}\\
\textit{Prime University}\\
Dhaka, Bangladesh.\\
ahmedalamin2357@gmail.com}

\and
\IEEEauthorblockN{Md Khairul Islam}
\IEEEauthorblockA{\textit{Dept. of Information and}\\ \textit{Communication Technology} \\
\textit{Islamic University}\\
Kushtia-7003, Bangladesh.\\
mdkito51@gmail.com}

\and
\IEEEauthorblockN{Bikash Chandra Singh}
\IEEEauthorblockA{\textit{Dept. of Information and}\\ \textit{Communication Technology} \\
\textit{Islamic University}\\
Kushtia-7003, Bangladesh.\\
bikash.singh@ice.iu.ac.bd}

\and
\IEEEauthorblockN{Md. Imran Hossain Showrov}
\IEEEauthorblockA{\textit{Institute of Computer Science,} \\
\textit{Bangladesh Atomic Energy}\\
\textit{ Commission,}\\
Dhaka, Bangladesh.\\
showrov@baec.gov.bd}

\and
\IEEEauthorblockN{Anirudda Sarkar}
\IEEEauthorblockA{\textit{Faculty of Mathematics} \\
\textit{and Computer Science}\\
\textit{South Asian University,}\\
New Delhi, India.\\
anirudda.cs@gmail.com}

}

\maketitle

\begin{abstract}
Sentiment analysis can largely influence the people to get the update of the current situation. Coronavirus (COVID-19) is a contagious illness caused by the coronavirus 2 that causes severe respiratory symptoms. The lives of millions have continued to be affected by this pandemic, several countries have resorted to a full lockdown. During this lockdown, people have taken social networks to express their emotions to find a way to calm themselves down. People are spreading their sentiments through microblogging websites as one of the most preventive steps of this disease is the socialization to gain people's awareness to stay home and keep their distance when they are outside home. Twitter is a popular online social media platform for exchanging ideas. People can post their different sentiments, which can be used to aware people. But, some people want to spread fake news to frighten the people. So, it is necessary to identify the positive, negative, and neutral thoughts so that the positive opinions can be delivered to the mass people for spreading awareness to the people. Moreover, a huge volume of data is floating on Twitter. So, it is also important to identify the context of the dataset. In this paper, we have analyzed the Twitter dataset for evaluating the sentiment using several machine learning algorithms. Later, we have found out the context learning of the dataset based on the sentiments. 
\end{abstract}


\section{Introduction}
Sentiment implies a view that is held or reflected in a statement or viewpoint. Analysis of sentiment (a type of text analytics) tests the behavioral intent towards the facets of an item mentioned in the text. Natural language processing (NLP) and machine learning methods are used in a textual sentiments analytics method to give weighted sentiment ratings to entities, topics, themes, and categories inside a phrase or sentence \cite{liu2010sentiment}. Sentiment analysis lets researchers of big business data gauge public sentiment, execute complex market polling, monitor brand and product image, and interpret consumer experiences. A clear method approaches basic sentiment analysis of text documents: 1) Break down every textual document into its constituent pieces (sentences, phrases, tokens, and parts of speech) 2) Analyze each sentiment-bearing phrase and aspect. 3) Assign a sentiment score to each phrase and component (from -1 to +1)\cite{liu2010sentiment}.

COVID-19 is an aggressive disease caused by a newly recognized coronavirus. The majority of people infected with COVID-19 experience mild to severe respiratory failure and survive without any additional treatment. In December 2019, a large number of individuals in Wuhan, China, suffered pneumonia due to an unexplained reason. Through contact monitoring, it was identified that these sufferers were related to the Wuhan seafood and wet animal wholesale industry\cite{world2020novel}. To explore the symptoms further, Chinese officials used deep sequencing analysis, which revealed enough proof that the new coronavirus, identified as COVID-19, was the causal agent of the sickness. Since then, COVID-19 has expanded all across China and the rest of the world..

According to the World Health Organization, one of the most effective ways to prevent this pandemic from spreading at such a rapid pace is through isolating and self-quarantine. Individual activity on social media sites like Facebook, Twitter, and YouTube tended to rise with the spread of the COVID-19 infection worldwide. Several studies have demonstrated that as a source of information, social media may play a crucial role in detecting epidemics. It also helps to understand public attitudes and actions during a crisis in order to promote crisis awareness and communication for health improvement. Twitter, one of the most popular microblogging sites, has become one of the key ways of sharing and self-documenting information. One of the forums for millions to share their feelings about various problems has been Twitter.

In this paper, we analyzed several countries' twitter sentiments about COVID-19 and also doing context analysis on user's sentiments (e.g. positive, negative, neutral). Finally, we built a model based on a machine learning approach to forecasting the thoughts based on their tweets about COVID19.

\section{Related Works}\label{RelatedWorks}
As the volume of generated data is increasing from different online sources, a significant number of researchers are involved in evaluating the social media data and gather available information. Recently, sentiment analysis catches the eye of researchers due to its substantial amount of work and rapid growth.

Online social media is a great platform to evaluate the emerging pattern of COVID-19. From the dataset, it can also be found that how the people dealing with this pandemic. The sentiment analysis, also known as opinion mining, employs natural language processing approaches to analyze a person's opinion and mood\cite{alsaeedi2019study}. To classify that sentiments, text mining, and deep learning techniques could be a common practice \cite{rout2018model}.

In \cite{kaur2021proposed}, Kaur et al.  presented a deep learning-based approach for categorizing tweets based on COVID-19. They also compared the performance of the suggested model to that of the Recurrent Neural Network and the Support Vector Machine, and observed that the proposed model had the highest accuracy.

Shamrat et al. in \cite{shamrat2021sentiment} conducted sentiment analysis on a collection of tweets concerning COVID-19 vaccinations. To classify sentiments, they utilized NLP and the KNN algorithm. They identify that Pfizer, Moderna, and AstraZeneca received 47.29\%, 46.16\%, and 40.08\ positive sentiments, respectively, after doing the analysis.

Aloufi et al. \cite{aloufi2018sentiment} executed a domain-specific sentiment analysis model employing several machine learning approaches. They worked on the football specific domain, used several features such as goal scoring, penalties, fouls, etc. Their work helped to assess the reactions of the fans at different moments during the match. First, they created a dataset on the reactions of the fans of football. Further, they used the dataset to construct a sentiments lexicon and then used a classifier to evaluate the tweets.
Inspiring from the ordinal regression, Saad and Jing \cite{saad2019twitter} experimented on a detailed analysis of tweets employing various machine learning approaches. The overall process associated with the dataset preprocessing, then extract features and finally, applysing ML methods to categorize the tweets. The observed results indicated that machine learning algorithms may identify ordinal regression with outstanding results. The Decision Tree was shown to have the highest accuracy, with a value of 91.81\%.

In the beginning, Xiang \cite{xiang2013china} studied the image of China by collecting thousands of tweets and analyzing them. The topics of tweets are also manually analyzed.  Xiao et al. \cite{chen2020country} utilized ML approaches to examine the situation in general of China collecting the large-scale twitter data. However, the research of aspect-level feelings, which is necessary to visualize the intricacy of the country's image, was overlooked in the process.

\begin{figure}[h]
	\centering
	\includegraphics[scale = .3]{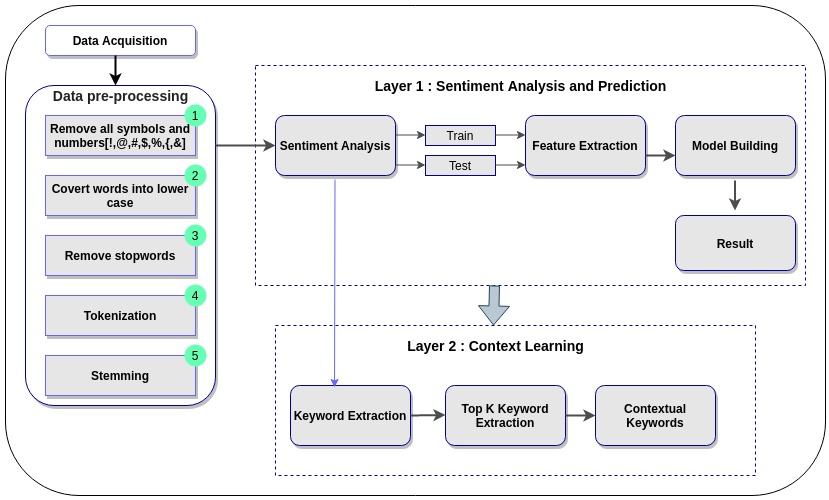}
	\caption{\centering A schematic model of proposed approach}\label{met}
\end{figure}

\begin{figure*}[h]

\begin{subfigure}{.5\textwidth}
  \centering
  \includegraphics[width=.7\linewidth]{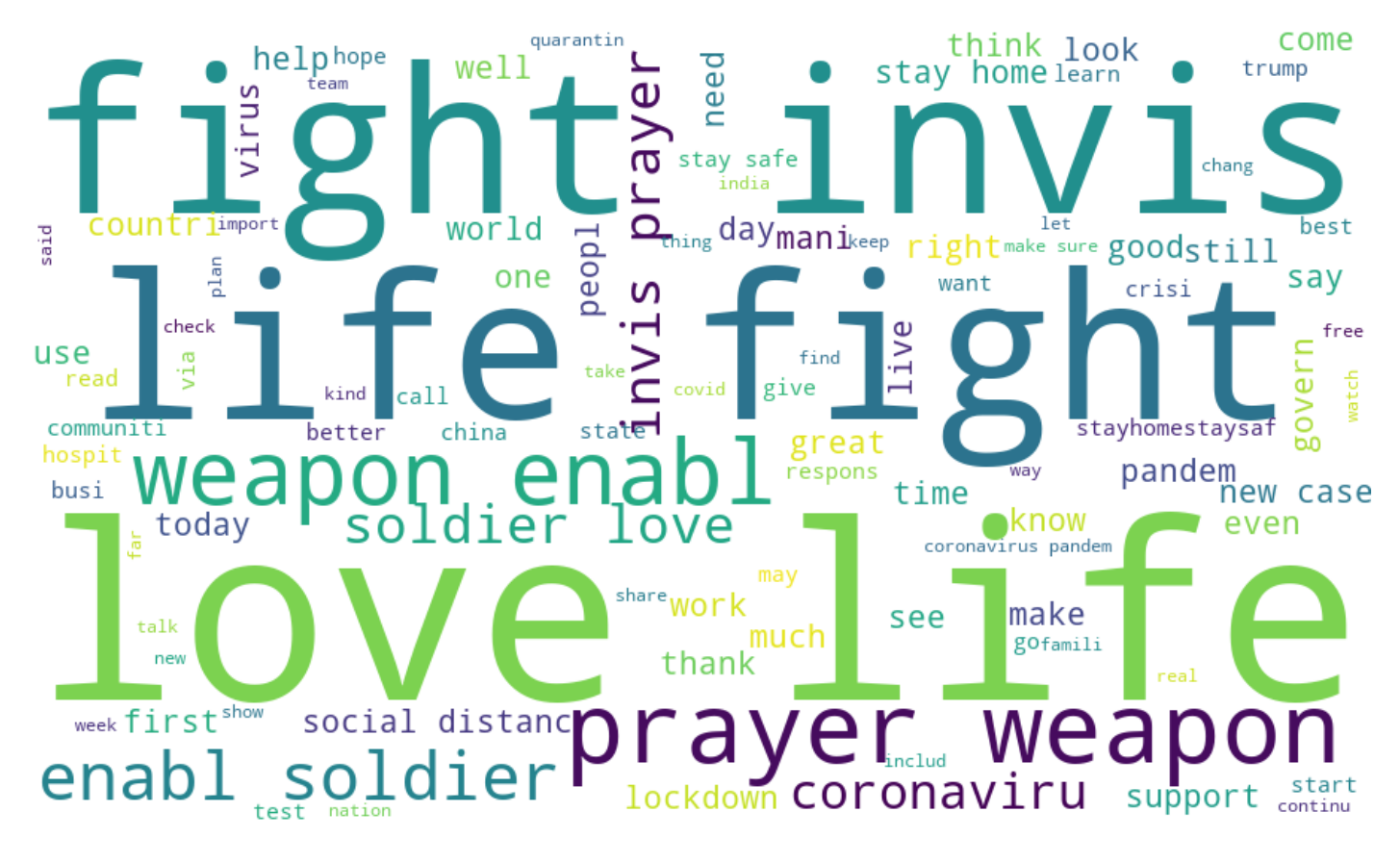}  
  \caption{Positive sentiments}
  \label{fig:sub-positive}
\end{subfigure}
\begin{subfigure}{.5\textwidth}
  \centering
  \includegraphics[width=.7\linewidth]{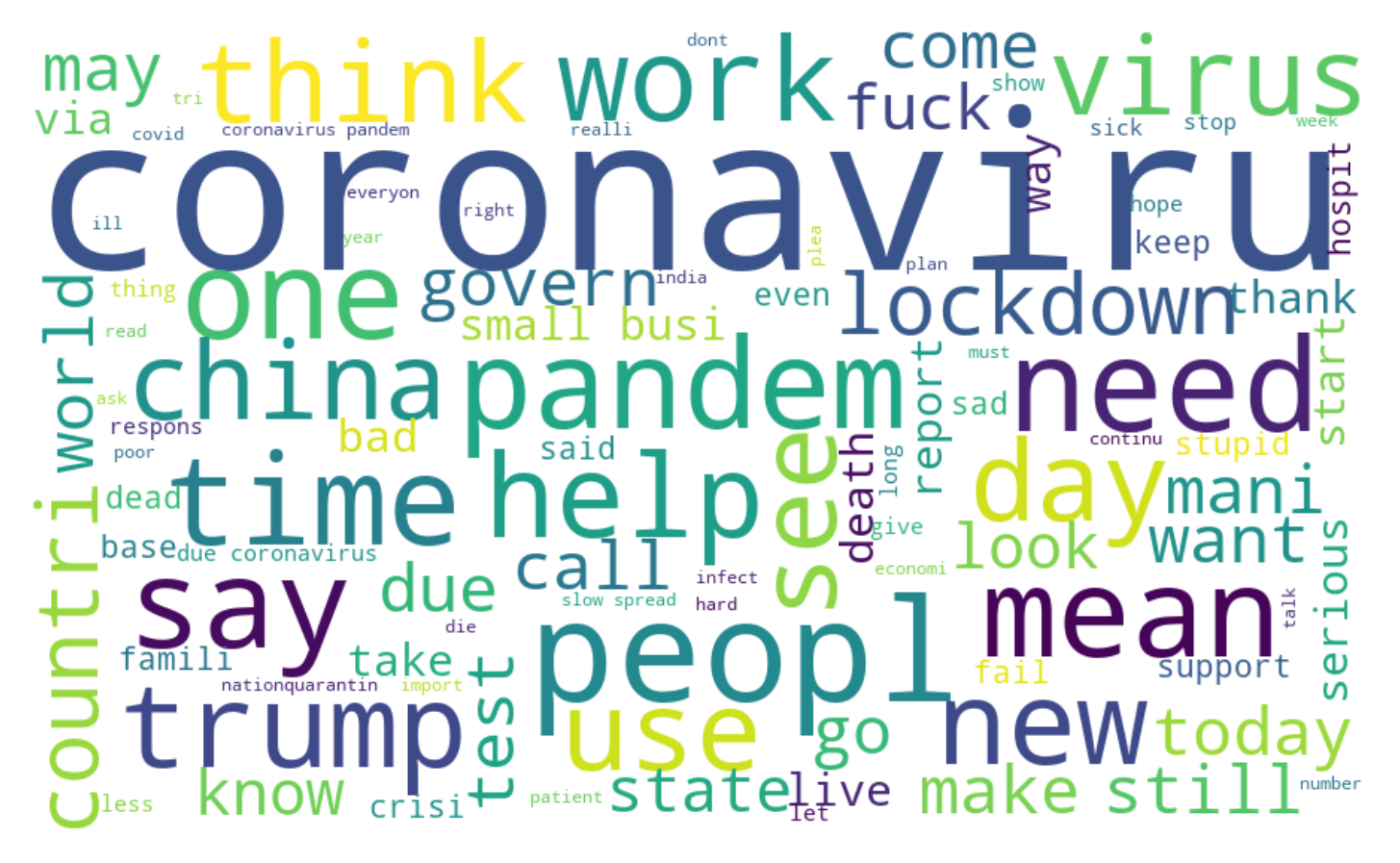}  
  \caption{Negative sentiments}
  \label{fig:sub-negative}
\end{subfigure}
\newline

\begin{subfigure}{.5\textwidth}
  \centering
  \includegraphics[width=.7\linewidth]{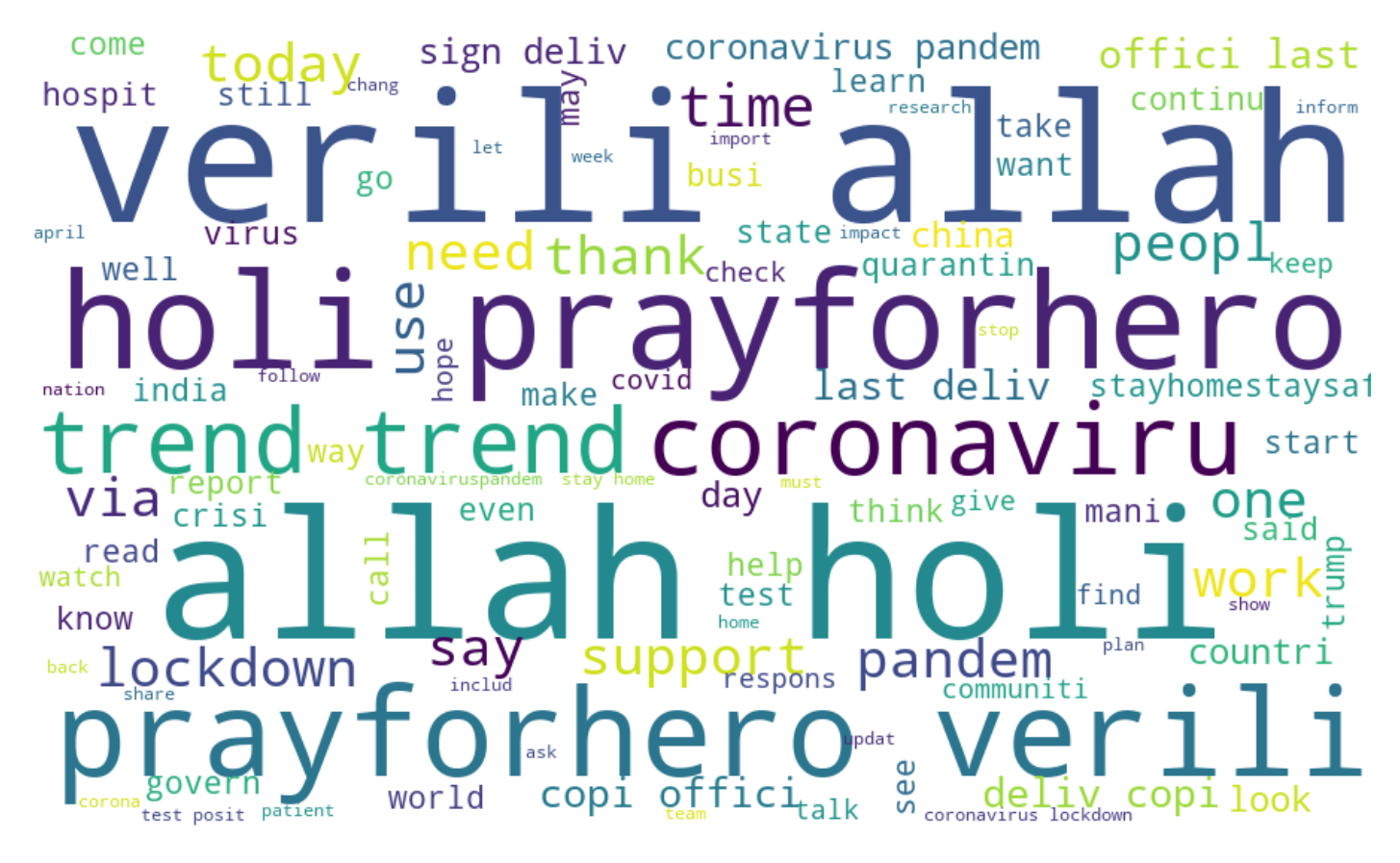}  
  \caption{Neutral sentiments}
  \label{fig:sub-neutral}
\end{subfigure}
\begin{subfigure}{.5\textwidth}
  \centering
  \includegraphics[width=.7\linewidth]{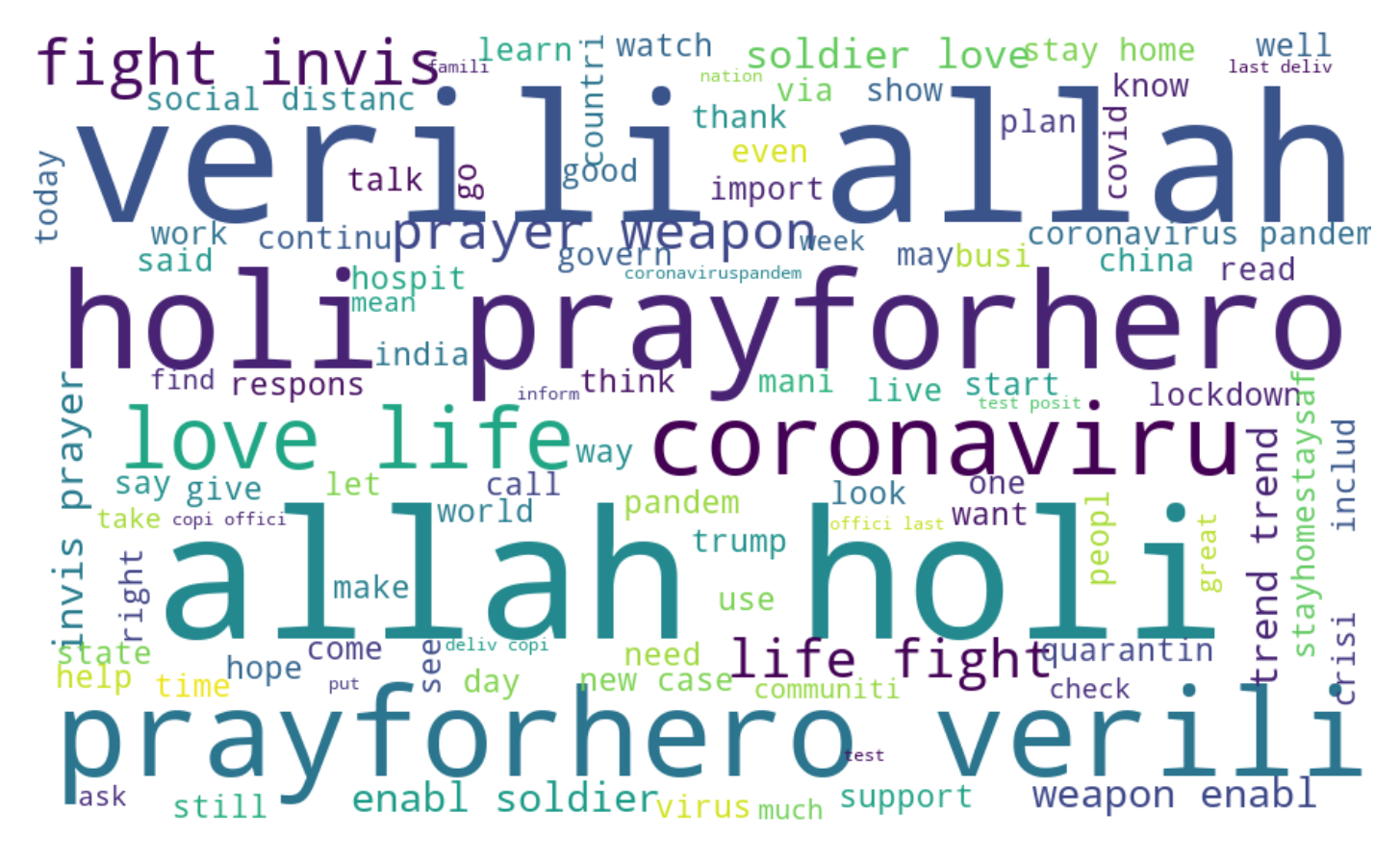}  
  \caption{Overall sentiments}
  \label{fig:sub-overall}
\end{subfigure}
\caption{Wordcloud of different sentiments of top-50 keywords obtained from dataset)}
\label{fig:sentiments}
\end{figure*}

\section{Methods}
This section illustrates the methodology of the proposed model. The overall process consists of two parts. Firstly, data acquisition and preprocessing  where data is collected from kaggle related to COVID-19. The second part is divided into two layers. In the first layer, we utilized a machine learning algorithm to assess sentiments from the dataset.. In the second layer, the extracted results are used to identify the context of the top-k keywords. All the steps carried out in this section are shown in \textbf{Figure \ref{met}}.
\subsection{Data acquisition and preprocessing}
Twitter is the primary source to collect tweets that are specific to the coronavirus disease. We collected coronavirus related tweets of several months in 2020 from the Kaggle \cite{dataset}.The dataset contains 357526 tweets regarding covid19. In order to preprocess the dataset, first, we removed non-English tweets.Then we took off punctuation (\#,!, \&, \%, \$), stop words like as, so, any, the so on, as well as non-printable characters like emojis. We simplified user references on Twitter by substituting them with the term username, such as "@Alaa" to "@username". Word Tokenization is performed further, which divides a chunk of text into distinct words depending on a specific separator\cite{webster1992tokenization}. Finally, we did stemming, which is typically accomplished by eliminating any associated suffixes and prefixes (affixes) from index words before assigning the phrase to the index\cite{jivani2011comparative}. We utilized regular expressions\cite{li2008regular} and NLTK(Natural Language Toolkit)\cite{bird2006nltk} to preprocess our data.

\subsection{Layer 1: sentiment analysis and prediction}
We implemented sentiment analysis\cite{alsaeedi2019study}, which takes the average amount of retweets, likes, and follows for each subject and then determines the engagement level for each. The Python textblob library\cite{loria2018textblob} was used to do sentiment assessment on the text corpus. The sentiment index ranged from –1.0 to 1.0, with –1.0 being the most negative content and 1.0 representing the most positive text. When we use an NLP algorithm, it always works with numbers. We are unable to send our text straight into that algorithm. As a result, bag-of-words and tf-idf are used to extract text characteristics by turning them into numeric vectors.

\subsubsection{Bag of words}
The major methodology offered by information extraction researchers to describe text corpus is bag of words\cite{el2016enhancement} presentation, which is a straightforward method for transforming raw data to structured data phrase by phrase while avoiding grammar.

\subsubsection{Tf-idf}
Term frequency–inverse document frequency(TF-IDF) is a numerical measure meant to indicate the importance of a term in a gathering or corpus of documents\cite{zhang2011comparative}.The frequency of a term(t) in a document(d) is measured by Term frequency(TF)\cite{zhang2011comparative}. The IDF is a measure for determining how significant a term(t) is \cite{zhang2011comparative}.

\begin{equation}
    TF(t,d) = \frac{\textrm{count of t in d}}{\textrm{number of words in d}}
\end{equation}

\begin{equation}
    IDF(t) = log{\frac{\textrm{number of documents}}{\textrm{number of documents with term 't'}}}
\end{equation}

\begin{equation}
    TF-IDF(t,d) = TF(t,d) * IDF(t)
\end{equation}

Finally, We will use several machine learning algorithms such as Logistic Regression, Naive Bayes, Decision Tree, Random Forest, XgBoost, Support Vector Machine  Algorithms to train our proposed model.

\subsubsection{Logistic Regression}
Logistic regression is the preferred regression analysis to use where the dependent variable is dichotomous (binary)\cite{pranckevivcius2017comparison}. Logistic regression is called after the logistic function, which is utilized at the heart of the approach. The logistic function, commonly known as the sigmoid, is an S-shaped curve that can take any real-valued integer and translate it to a range between 0 and 1, but never precisely at those boundaries\cite{pranckevivcius2017comparison}.

\subsubsection{Support Vector Machine}
The Support Vector Machine (SVM) is a plane-based classifier that creates a discrete hyperplane in the training data and compounds' descriptor space \cite{pranckevivcius2017comparison}. By solving an optimization problem, SVM finds the optimal hyperplane between two groups of data in the training data \cite{pranckevivcius2017comparison}.

\subsubsection{Naive Bayes}
Naive Bayes is a categorization approach based on Bayes' Principle and the assumption of predictor independence. \cite{pranckevivcius2017comparison}. Bayes theorem gives a way to calculate posterior probability from prior probability of class, prior probability of predictor and likelihood which is the probability of predictor.

\subsubsection{Decision Tree}
Decision Tree addresses problems by employing a tree representation where each leaf node corresponds to a class labels and characteristics are specified on the interior node of the tree. \cite{pranckevivcius2017comparison}.Based on the provided attributes, information gain calculates the difference between the dataset's entropy before and after splitting.

\subsubsection{Random Forest}
Random Forest is a group of classification or regression trees that can acquire equivalent training datasets, known as bootstraps, and then merge them to get a more reliable result \cite{pranckevivcius2017comparison}.

\subsubsection{XGBoost}
XGBoost is a flexible and cutting-edge implementation of gradient boosting machines that have been shown to exceed the computing power limits for enhanced tree methods\cite{chen2015xgboost}.


\begin{table*}[h]
\centering
\caption{Evaluation results of sentiment analysis over different machine learning model using the bag-of-words and tf-idf as feature extractor}\label{tab:accuracy_table}
{
\begin{tabular}{|c|l|c|c|c|c|}
\hline

 Feature Extraction  & Algorithms & Accuracy & Precision & Recall & $F_1$-score \\
\hline
\multirow{7}{*}{Bag-of-words}
&Logistic Regression & 92.55\% & 93.18\% & 87.90\% & 90.07\% \\

&SVM & 92.77\% & 93.45\% & 88.27\% & 90.40\% \\
&Naïve Bayes & 84.71\% & 80.16\% & 82.76\% & 80.99\% \\

&Decision Tree & 89.17\% & 86.33\% & 86.02\% & 86.17\% \\
&Random Forest & \textbf{93.00\%} & 93.3\% & 88.92\% & 90.78\% \\
&XGBoost & 80.56\% & 86.63\% & 68.7\% & 72.18\% \\

\hline
\multirow{7}{*}{TF-IDF}
& Logistic Regression & 88.96\%& 90.78\% & 81.09\% & 84.22\%\\
&SVM & 92.18\% & 92.92\% & 87.18\% & 89.48\% \\
& Naïve Bayes & 88.79\% & 88.87\% & 81.93\% & 84.30\%\\
&Decision Tree & 89.20\% & 86.46\% & 85.67\% & 86.05\%\\
& Random Forest & \textbf{92.97\%} & 93.37\% & 88.70\% & 90.66\%\\
&XGBoost & 80.35\% & 86.54\% & 68.50\% & 71.98\% \\

\hline
\end{tabular}
}
\end{table*}

\subsection{Layer 2: Context Learning}
In the second layer, we apply keyword ranking algorithm to find out the importance value of the keywords within the text. After evaluating keyword, we rank them according to their importance value. Then we select $top-k$ keywords (for our case, $k = 50$) for the context evaluation purpose. 

\section{Experiment Result Analysis}\label{Result_and_Limitation}
In the first experiment, we find the amount of each label of sentiment classes in percentage. The next experiment shows the accuracy of the machine learning approaches on testing dataset. And we find the best machine learning model to predict the sentiment according to the accuracy.

\subsection{Experimental Setup and Parameters Setting}
We utilized the scikit-learn(version: 0.22.2) library's CountVectorizer and TfidfVectorizer tools in Python(version: 3.7) to conduct bag-of-words and tf-idf on processed text data. We employed Python's scikit-learn library to develop a machine learning model. When utilizing the tf-idf and the bag-of-words methods as feature extractors, we excluded terms that appeared fewer than 500 times in the texts.

\subsection{Sentiment Analysis}
\begin{figure}[htb!]
	\centering
	\includegraphics[scale=.22]{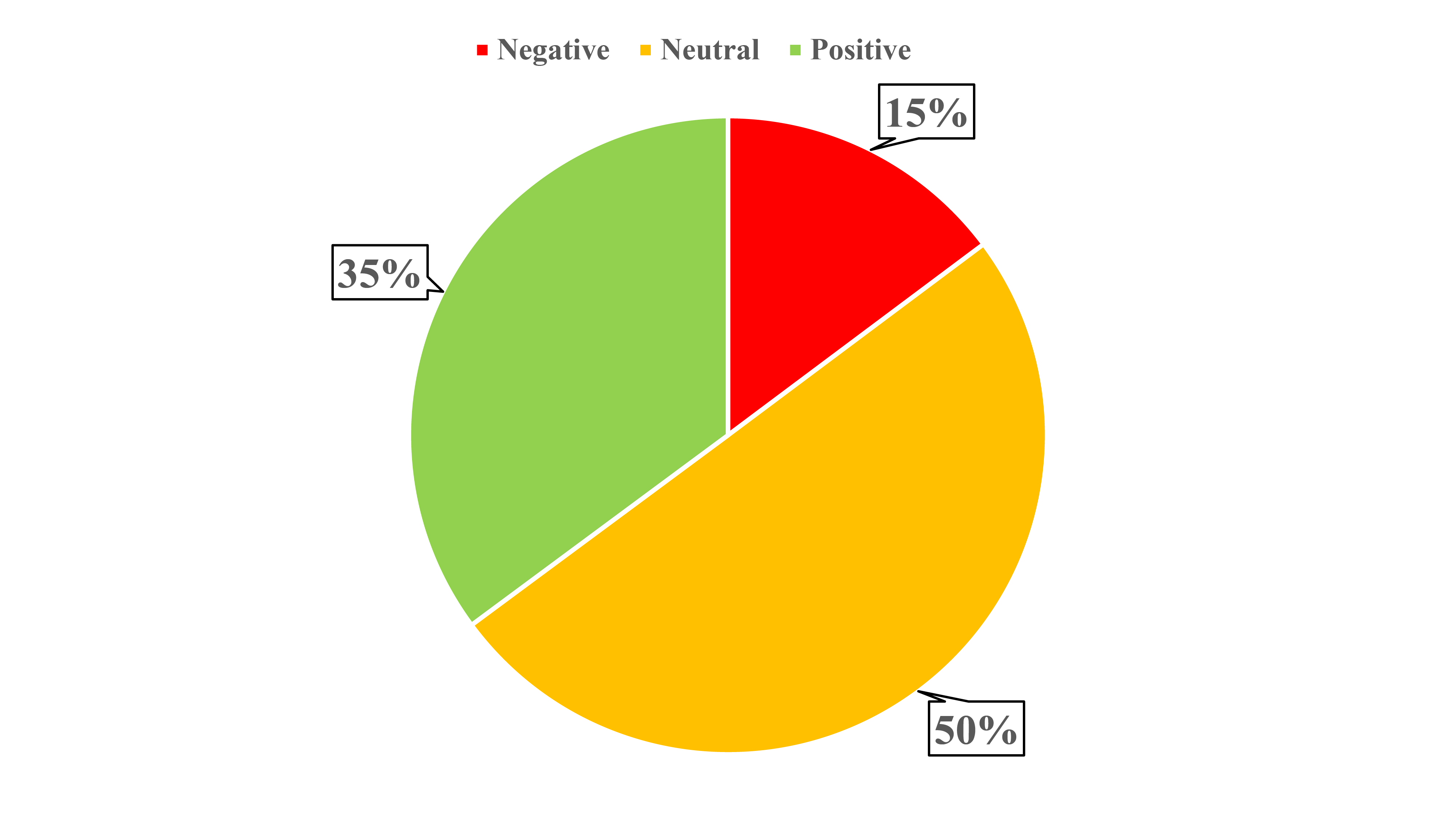}
	\caption{\centering Sentiment analysis results}\label{sentiment_result}
\end{figure}

We have collected covid-19 twitter dataset from the Kaggle repository. It contains 357526 sentiments of people regarding the covid-19 situation. There, after evaluating the sentiment,  we have 179036 neutral tweets, 125634 tweets are positive and only 52856 tweets are negative. The percentage of each sentiment is illustrated in \textbf{Figure \ref{sentiment_result}}, where 50\% of sentiments are neutral, 35\% of sentiments are positive and just 15\% of sentiments are negative.

\subsection{Evaluation Matrix}
We considered four standard metrics: accuracy, precision, recall, and the f1-score. Accuracy (A) refers to the number of valid predictions divided by the total number of input samples. Precision(P) is calculated by dividing the number of right positive results by the number of positive outcomes predicted by the classifier. Recall (R) is the number of correct certain outcomes divided by the total number of conjugate samples. F1-score (F1) is also known as harmonic mean, and it seeks to strike a compromise between precision and recall. It computes with both false positives and false negatives and works well on an imbalanced dataset. Finally, the AUC-ROC curve\cite{narkhede2018understanding} is calculated to evaluate the performance.

\subsection{Result Analysis}
We constructed a model that distinguishes among sentiments in different labels (positive, neutral, and negative) using the collected dataset. Firstly, we eliminated ambiguity from the dataset by preprocessing it. After preprocessing, we separated our dataset into two parts, with 80\% used to train our machine learning models and the remaining 20\% used to test our models.Then, to extract features from the text, we used  bag of words and tf-idf methods. Machine learning classification models have been used after transforming characters into numerical vectors.

\begin{figure*}[h]

\begin{subfigure}{.5\textwidth}
  \centering
  \includegraphics[width=.9\linewidth]{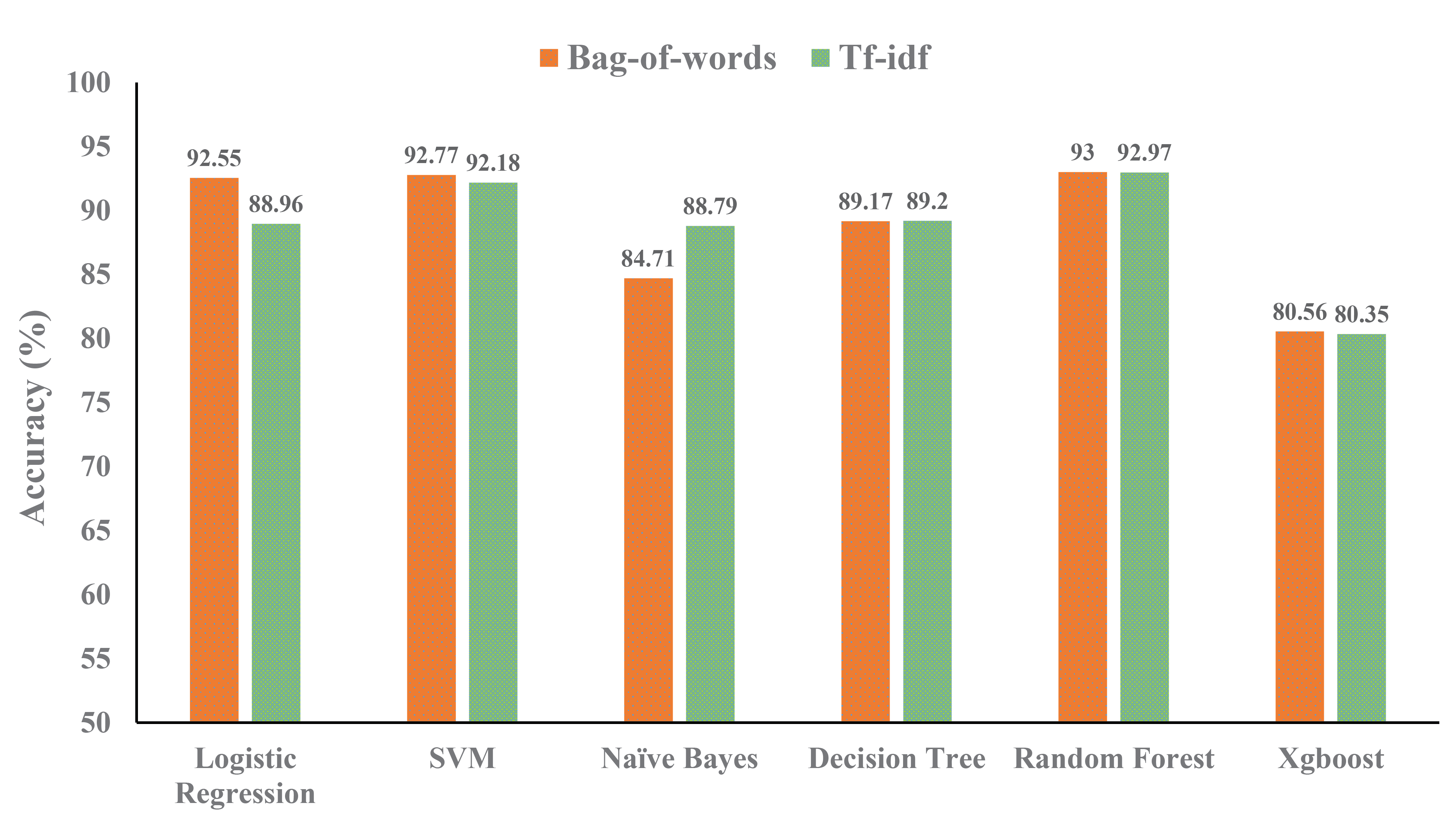}  
  \caption{Accuracy}
  \label{fig:sub-accuracy}
\end{subfigure}
\begin{subfigure}{.5\textwidth}
  \centering
  \includegraphics[width=.9\linewidth]{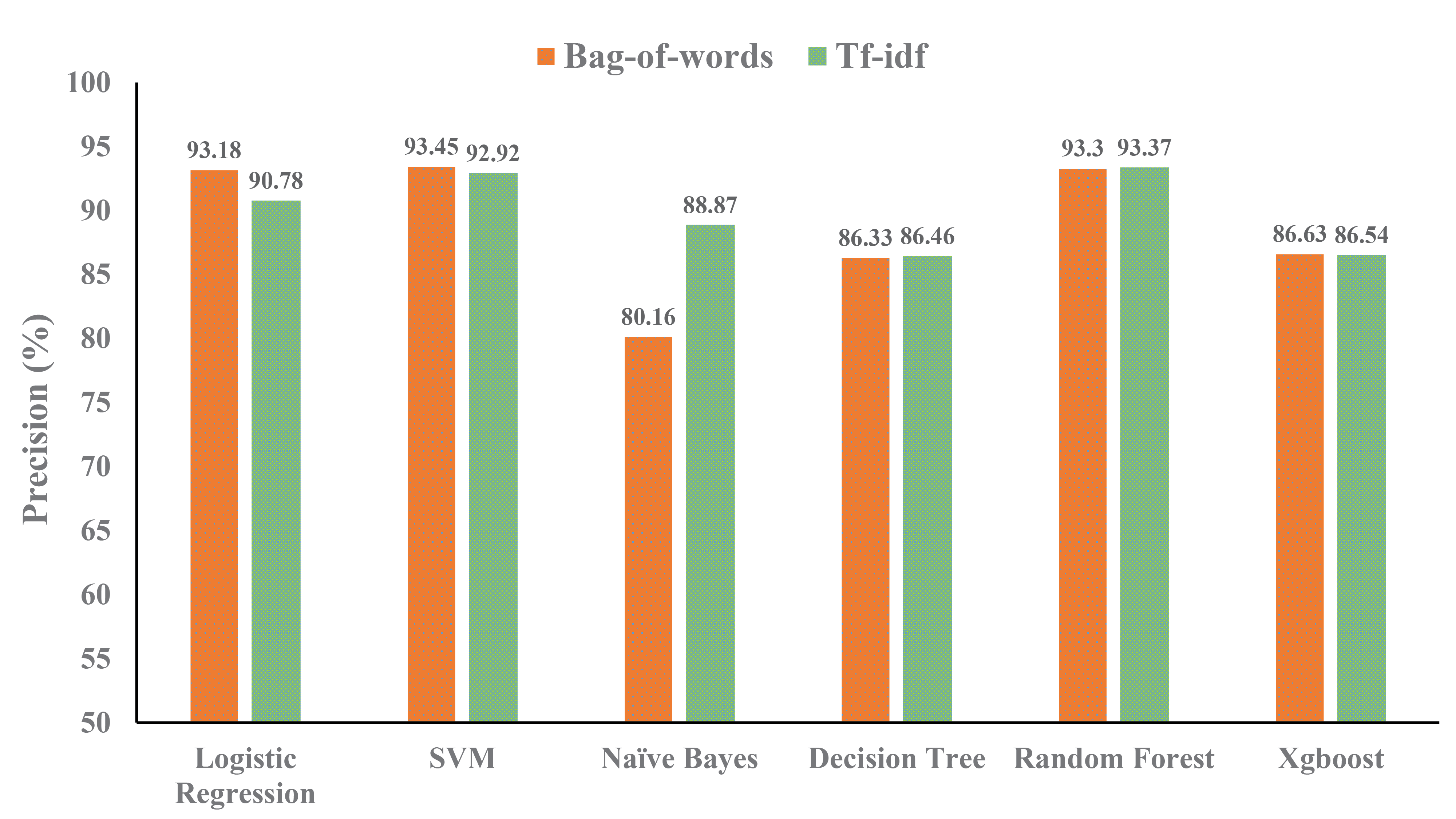}  
  \caption{Precision}
  \label{fig:sub-precision}
\end{subfigure}
\newline

\begin{subfigure}{.5\textwidth}
  \centering
  \includegraphics[width=.92\linewidth]{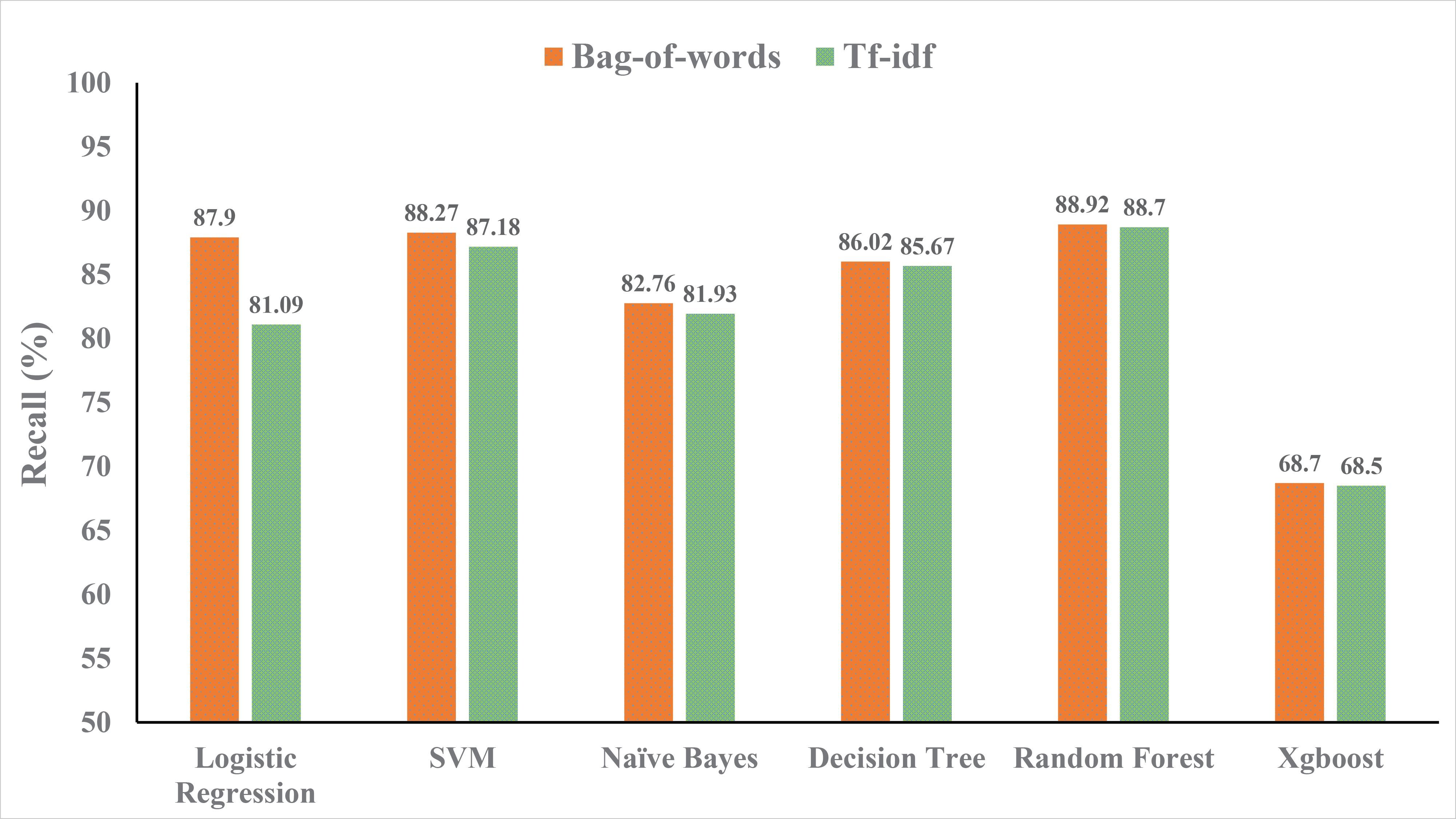}  
  \caption{Recall}
  \label{fig:sub-recall}
\end{subfigure}
\begin{subfigure}{.5\textwidth}
  \centering
  \includegraphics[width=.92\linewidth]{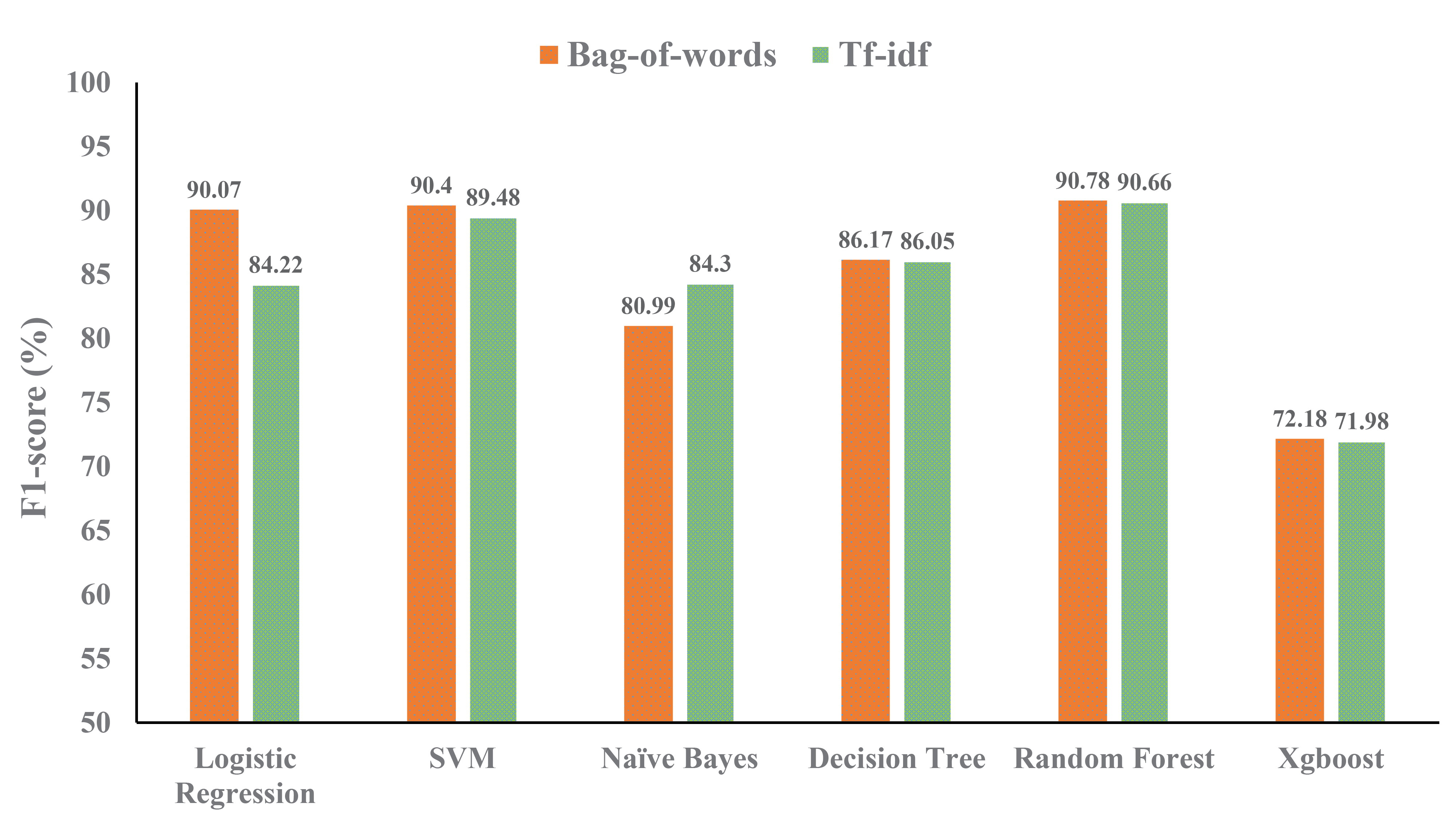}  
  \caption{F1-score}
  \label{fig:sub-f1}
\end{subfigure}
\caption{Performance comparison results (accuracy, precision, recall, and f1-score)}
\label{fig:fig-performance}
\end{figure*}

\textbf{Table \ref{tab:accuracy_table}} demonstrates the performance obtained for each classification algorithms. The performances of logistic regression are 92.55\% for bag of words and 88.96 \% for tf-idf , SVM provides 92.77\% for bag of words and 92.18\% for tf-idf, Naive Bayes provides 84.71\% for bag of words and 88.79\% for tf-idf, Decision Tree provides 89.17\% for bag of words and 89.2\% for tf-idf , the random forest provides 93\% for bag of words and 92.97\% and Xgboost provides 80.56\% for bag of words and 80.35\% for tf-idf in terms of accuracy. Where the accuracy of naive Bayes, decision tree, xgboost, svm, logistic regression is less than 93\%. Among all of the models, the random forest provides the highest accuracy for both bag of words and tf-idf.  As we are also interested in how well our model classifies each sentiment, we also find the other performance metrics like precision, recall, and f1-score of each model. In this case, the f1-score of the random forest again performed well with 90.78\% for bag of words and 90.66\% for tf-idf. And again, naive Bayes, decision trees, and xgboost perform comparatively low. \textbf{Figure \ref{fig:fig-performance}} depicts a visual comparison of each machine learning method for bag of words and tf-idf.

From the analysis, random forests can distinguish true positive and true negative more accurately than other models. There are several causes for the result. For multi-class concerns, Random Forest is inherently appropriate. As the proposed model is a multiclass based problem, so with a combination of numerical and categorical characteristics, Random Forest operates well. Random Forest provides the possibility of class labels for a multi-class classification task. Random forests function well when the dataset has a non-linear pattern and inference outside of the training data is not relevant.

The Xgboost is the model that got the worst result in this research. When data includes a combination of numerical and category information, or solely numerical features, Xgboost performs well. It doesn't provide good performance for the  problems with natural language processing and comprehension. Another prerequisite for Xgboost to function is that regression tasks that involve predicting a continuous output.

\begin{table*}[ht]
	\centering
	\caption{Top-50 keywords for different sentiments}\label{result_Keywords}
	\renewcommand{\arraystretch}{1.3}
	\begin{tabular}{|p{15mm}|p{70mm}|p{70mm}|} 
		\hline
		\textbf{Sentiment Type} & \textbf{Description} & \textbf{Top-50 Keywords}\\
		\hline        				
		Positive & People talk about helping each other, social awareness, and supporting each other in positive sentiment. They talk about staying at home, social distancing, and figuring out how covid19 will stop spreading. They have showered their gratitude and praised doctors, nurses, scientists, soldiers serving outdoors and delivering emergency services to the general public in this pandemic & coronavirus, love, fight, life, prayer, new, enabl, weapon, invis, soldier, people, case, live, time, pandem, need, help, get, work, good, lockdown, death, one, test, home, stay, right, safe,	like, health, make, great, day, first, thank, today, social, support, take, keep, report, say, word, free, latest, countri, much, see, care, know.
		\\ \hline
		
		Negative & In a pessimistic sense, most persons talk about their families' deaths and not receiving life-saving medical attention while they need it. During the coronavirus crisis, they worry about workers who are extremely vulnerable. People are talking about saving their tiny company affected by COVID-19. The school absolutely shut down due to COVID-19 is also a negative sentiment among most tweeted concerns & coronavirus, people, due, time, get, help, pandem, test, lockdown, need, like, work, case, death, health, one, busi, home, day, trump, say, week, make, spread, know, die, mean, take, even, virus, countri, govern, small, mani, least, other, want, support, new, world, care, dead, state, china, hard, see, think, use, would, sick.			
		
		\\ \hline
		
		Neutral	& People often speak, in a neutral way, about prayers to the Almighty. They share several health care tips that can be obtained during times of crisis to effectively care for workers. They post several videos and blogs that explain how to correctly wash your hands and regular self-care. People are often sharing several donations to help fight the virus & coronavirus, allah, prayforhero, holi, verili, trend, pandem, lockdown, people, help, test, support, case, need, get, time, work, health, deliy, death, like, day, home, say, trump, one, fight, tody, via, thank, last, world, state, virus, china, crisi, use, govern, respons, stayhomestaysafe, take, covid, make, pleas, countri, sign, worker, stayhom, know, offici.
		
		\\ \hline 	
		Overall & Overall sentiment of the people & coronavirus, allah, prayforhero, holi, verili, people, fight, pandem, love, life, help, case, lockdown, time, new, test, need, get, prayer, enable, weapon, trend , work, soldier, inyis, death, like , health, support, one, home, day, make, say, live, today, thank, take, world, trump, stay, virus, state, countri, know, use, busi, crisi, care.
		
		\\ \hline
	\end{tabular}
	\vspace{1ex}	
\end{table*}

\begin{figure}[h]
	\centering
	\includegraphics[width=8.5cm]{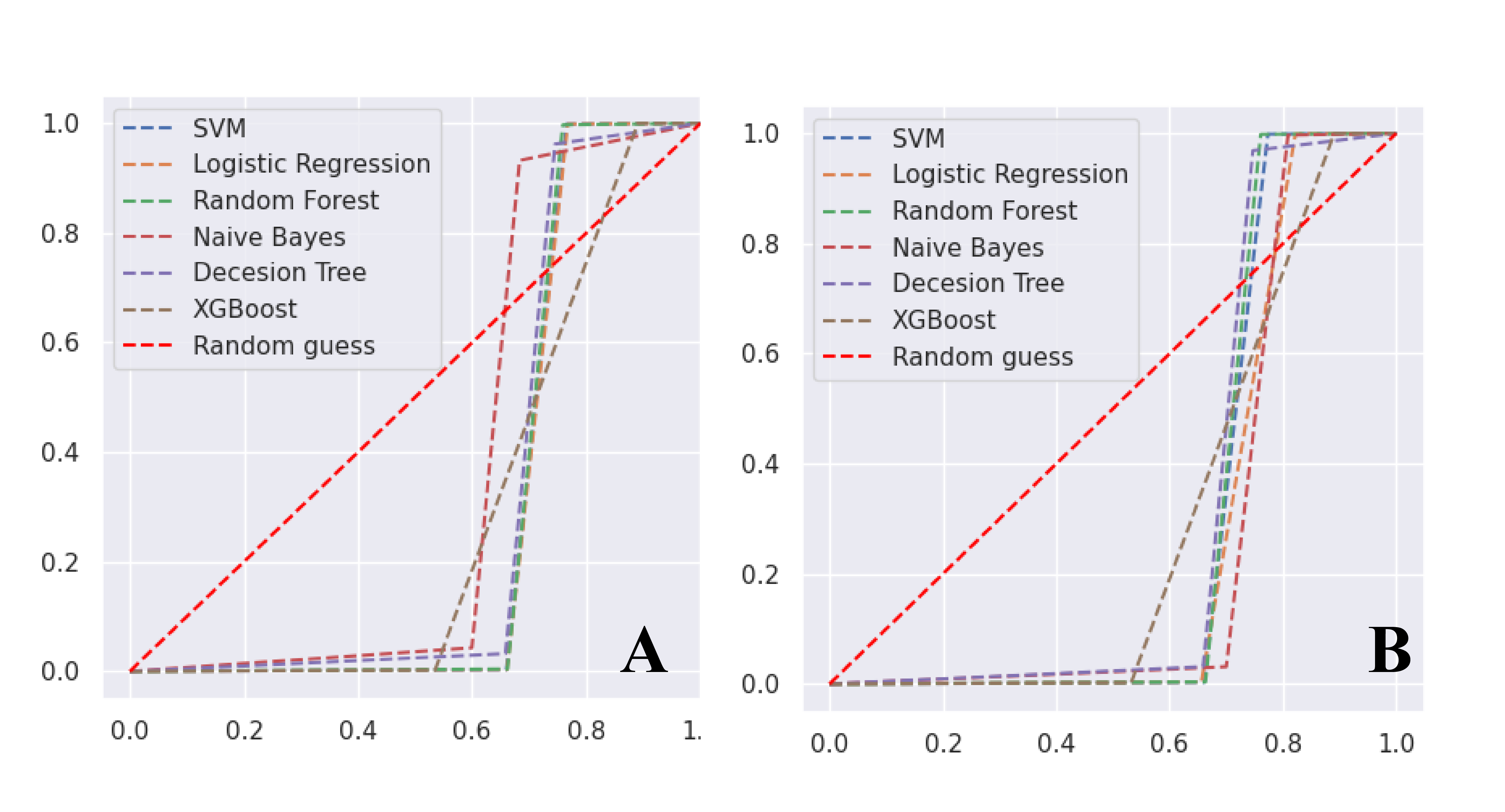}
	\caption{Comparing ROC-AUC Curve. \textbf{(A)} ROC-AUC Curve for Bag-of-words. \textbf{(B)} ROC-AUC Curve for TF-IDF.}\label{fig5}
\end{figure}

\textbf{Figure \ref{fig5}} illustrates the ROC curve and AUC values of all models where random forest provides the highest AUC value among all the models which is 0.90. Xgboost provides the lowest AUC value which is 0.702. Since random forest provides the highest auction benefit, it provides better output for all models to differentiate between the positive, negative, and neutral classes.

Finally, we have performed context learning on the dataset to identify the most frequent words in each sentiment. We carried out keyword extraction on each sentiment after identifying the sentiment to determine the top-50 frequent words. We have the top-50 positive words from the positive sentiment, the top-50 negative sentiment words, and the top-50 neutral words from the neutral sentiment. The top-50 frequent words of each sentiment are shown in \textbf{Table \ref{result_Keywords}}. The obtained result of context learning shows that positive sentiments give 80\% accuracy, negative sentiments give 84\% and neutral sentiments give 78\% accuracy. All the accuracy is calculated based on \textit{top-50} keywords. 

\textbf{Figure \ref{fig:sentiments}} depicts a wordcloud of the most frequently occurring words in the dataset. \textbf{Figure \ref{fig:sub-positive}} indicates that the most often used words in positive sentiment are \textit{love} and \textit{life}. \textbf{Figure \ref{fig:sub-negative}} reveals that \textit{people}, \textit{due} appear most frequently in negative sentiment. \textbf{Figure \ref{fig:sub-neutral}} and \textbf{Figure \ref{fig:sub-overall}} indicate the wordcolud for the neutral sentiment and overall sentiments, respectively.

\begin{table}[ht]
	\centering
	\caption{Evaluation results of context learning based on the top-50 keywords for different sentiments}\label{result2}
	\renewcommand{\arraystretch}{1.3}
	\begin{tabular}{ll}
		\hline
		\textbf{Sentiment Type} & \textbf{Accuracy} \\		\hline        				
		Positive & 80\% \\ \hline
		Negative & 84\%\\ \hline
		Neutral & 78\%  \\ \hline
	\end{tabular}
	\vspace{1ex}	
\end{table}

\section{Conclusion and Future work}\label{conclusion_and_future_work}
The purpose of this research was to examine people's thoughts and emotions during the COVID-19 epidemic. The COVID-19 outbreak has wreaked havoc on numerous medical systems and governments, taking countless lives. People are spreading their thoughts through the social media platform as they are suggested to stay home and maintain social distancing. As a result, social media like microblogging websites are one of the major sources of sentiment related data. People are spreading their emotions that might be helpful to make aware to others. But there are chances to spread the fake news as well. We examined the sentiment using different machine learning approaches and then determined the context of the dataset in the proposed model. In the future, this analysis can be utilized to evaluate the evolving feelings and sentiments of individuals in various nations and see if there are any significant changes over time.

\bibliographystyle{IEEEtran}
\bibliography{referenceIEEE}

\begin{thebibliography}{10}
\providecommand{\url}[1]{#1}
\csname url@samestyle\endcsname
\providecommand{\newblock}{\relax}
\providecommand{\bibinfo}[2]{#2}
\providecommand{\BIBentrySTDinterwordspacing}{\spaceskip=0pt\relax}
\providecommand{\BIBentryALTinterwordstretchfactor}{4}
\providecommand{\BIBentryALTinterwordspacing}{\spaceskip=\fontdimen2\font plus
\BIBentryALTinterwordstretchfactor\fontdimen3\font minus
  \fontdimen4\font\relax}
\providecommand{\BIBforeignlanguage}[2]{{%
\expandafter\ifx\csname l@#1\endcsname\relax
\typeout{** WARNING: IEEEtran.bst: No hyphenation pattern has been}%
\typeout{** loaded for the language `#1'. Using the pattern for}%
\typeout{** the default language instead.}%
\else
\language=\csname l@#1\endcsname
\fi
#2}}
\providecommand{\BIBdecl}{\relax}
\BIBdecl

\bibitem{liu2010sentiment}
B.~Liu \emph{et~al.}, ``Sentiment analysis and subjectivity.'' \emph{Handbook
  of natural language processing}, vol.~2, no. 2010, pp. 627--666, 2010.

\bibitem{world2020novel}
WHO, ``World health organization (2020): Novel coronavirus (2019-ncov),
  situation report, 3,'' 2020.

\bibitem{alsaeedi2019study}
A.~Alsaeedi and M.~Z. Khan, ``A study on sentiment analysis techniques of
  twitter data,'' \emph{International Journal of Advanced Computer Science and
  Applications}, vol.~10, no.~2, pp. 361--374, 2019.

\bibitem{rout2018model}
J.~K. Rout, K.-K.~R. Choo, A.~K. Dash, S.~Bakshi, S.~K. Jena, and K.~L.
  Williams, ``A model for sentiment and emotion analysis of unstructured social
  media text,'' \emph{Electronic Commerce Research}, vol.~18, no.~1, pp.
  181--199, 2018.

\bibitem{kaur2021proposed}
H.~Kaur, S.~U. Ahsaan, B.~Alankar, and V.~Chang, ``A proposed sentiment
  analysis deep learning algorithm for analyzing covid-19 tweets,''
  \emph{Information Systems Frontiers}, pp. 1--13, 2021.

\bibitem{shamrat2021sentiment}
F.~J.~M. Shamrat, S.~Chakraborty, M.~Imran, J.~N. Muna, M.~M. Billah, P.~Das,
  and M.~O. Rahman, ``Sentiment analysis on twitter tweets about covid-19
  vaccines using nlp and supervised knn classification algorithm,''
  \emph{Indonesian Journal of Electrical Engineering and Computer Science},
  vol.~23, no.~1, pp. 463--470, 2021.

\bibitem{aloufi2018sentiment}
S.~Aloufi and A.~El~Saddik, ``Sentiment identification in football-specific
  tweets,'' \emph{IEEE Access}, vol.~6, pp. 78\,609--78\,621, 2018.

\bibitem{saad2019twitter}
S.~E. Saad and J.~Yang, ``Twitter sentiment analysis based on ordinal
  regression,'' \emph{IEEE Access}, vol.~7, pp. 163\,677--163\,685, 2019.

\bibitem{xiang2013china}
D.~Xiang, ``China's image on international english language social media,''
  \emph{Journal of International Communication}, vol.~19, no.~2, pp. 252--271,
  2013.

\bibitem{chen2020country}
H.~Chen, Z.~Zhu, F.~Qi, Y.~Ye, Z.~Liu, M.~Sun, and J.~Jin, ``Country image in
  covid-19 pandemic: A case study of china,'' \emph{IEEE Transactions on Big
  Data}, 2020.

\bibitem{dataset}
``https://www.kaggle.com/.''

\bibitem{webster1992tokenization}
J.~J. Webster and C.~Kit, ``Tokenization as the initial phase in nlp,'' in
  \emph{COLING 1992 Volume 4: The 14th International Conference on
  Computational Linguistics}, 1992.

\bibitem{jivani2011comparative}
A.~G. Jivani \emph{et~al.}, ``A comparative study of stemming algorithms,''
  \emph{Int. J. Comp. Tech. Appl}, vol.~2, no.~6, pp. 1930--1938, 2011.

\bibitem{li2008regular}
Y.~Li, R.~Krishnamurthy, S.~Raghavan, S.~Vaithyanathan, and H.~Jagadish,
  ``Regular expression learning for information extraction,'' in
  \emph{Proceedings of the 2008 Conference on Empirical Methods in Natural
  Language Processing}, 2008, pp. 21--30.

\bibitem{bird2006nltk}
S.~Bird, ``Nltk: the natural language toolkit,'' in \emph{Proceedings of the
  COLING/ACL 2006 Interactive Presentation Sessions}, 2006, pp. 69--72.

\bibitem{loria2018textblob}
S.~Loria, ``textblob documentation,'' \emph{Release 0.15}, vol.~2, p. 269,
  2018.

\bibitem{el2016enhancement}
D.~M. El-Din, ``Enhancement bag-of-words model for solving the challenges of
  sentiment analysis,'' \emph{International Journal of Advanced Computer
  Science and Applications}, vol.~7, no.~1, 2016.

\bibitem{zhang2011comparative}
W.~Zhang, T.~Yoshida, and X.~Tang, ``A comparative study of tf* idf, lsi and
  multi-words for text classification,'' \emph{Expert Systems with
  Applications}, vol.~38, no.~3, pp. 2758--2765, 2011.

\bibitem{pranckevivcius2017comparison}
T.~Pranckevi{\v{c}}ius and V.~Marcinkevi{\v{c}}ius, ``Comparison of naive
  bayes, random forest, decision tree, support vector machines, and logistic
  regression classifiers for text reviews classification,'' \emph{Baltic
  Journal of Modern Computing}, vol.~5, no.~2, p. 221, 2017.

\bibitem{chen2015xgboost}
T.~Chen, T.~He, M.~Benesty, V.~Khotilovich, Y.~Tang, H.~Cho \emph{et~al.},
  ``Xgboost: extreme gradient boosting,'' \emph{R package version 0.4-2},
  vol.~1, no.~4, pp. 1--4, 2015.

\bibitem{narkhede2018understanding}
S.~Narkhede, ``Understanding auc-roc curve,'' \emph{Towards Data Science},
  vol.~26, pp. 220--227, 2018.

\end{thebibliography}
\end{document}